\documentstyle[11pt,aaspp4]{article}
\begin{document}
%
%
%
%%%%%%%%%%%%%%%%%%%%%%%%%%%%%%%%%%%%%%%%%%%%%%%%%%%%%%%%%%%%%%%%%%%
%%%%%%%%%    This is the revised manuscript to       %%%%%%%%%%%%%%
%%%%%%%%%         appear in ApJ (pre-print version)  %%%%%%%%%%%%%%
%%%%%%%%%%%%%%%%%%%%%%%%%%%%%%%%%%%%%%%%%%%%%%%%%%%%%%%%%%%%%%%%%%%
%
%
%
%
%
%%%%%%%%%%%%%%%%%%%%%%%%%%%%%%%%%%%%%%%%%%%%%%%%%%%%%%%%%%%%%%%%%%%
%%%%%%%%%%%%%%%%%%%%%%%%%%%%%%%%%%%%%%%%%%%%%%%%%%%%%%%%%%%%%%%%%%%
\title{\huge Quasi Non-linear Evolution of Stochastic Bias}
%%%%%%%%%%%%%%%%%%%%%%%%%%%%%%%%%%%%%%%%%%%%%%%%%%%%%%%%%%%%%%%%%%%
%%%%%%%%%%%%%%%%%%%%%%%%%%%%%%%%%%%%%%%%%%%%%%%%%%%%%%%%%%%%%%%%%%%
%
%
\author{Atsushi Taruya$^1$, Kazuya Koyama$^2$ and Jiro Soda$^1$} 
\affil{
$1$, Department of Fundamental Sciences, FIHS, Kyoto University, 
Kyoto, 606-8501, Japan \\
$2$, Graduate School of Human and Environment Studies, Kyoto University, 
Kyoto, 606-8501, Japan 
}
\authoraddr{}
%
%
%
%
%
%
%
%
%
%%%%%%%%%%%%%%%%%%%%%%%%%%%definition%%%%%%%%%%%%%%%%%%%%%%%%%%%%%%%%%
\def\k{\mbox{\boldmath$k$}}
\def\x{\mbox{\boldmath$x$}}
\def\v{\mbox{\boldmath$v$}}
\def\dm{\delta_m}
\def\dg{\delta_g}
\def\Dm{\Delta_m}
\def\Dg{\Delta_g}
%%%%%%%%%%%%%%%%%%%%%%%%%%%%%%%%%%%%%%%%%%%%%%%%%%%%%%%%%%%%%%%%%%%%%%
%
%
%
%
\begin{abstract}
It is generally believed that the spatial distribution of galaxies does not 
trace that of the total mass. The understanding of the bias effect 
is therefore 
necessary to determine the cosmological parameters and the 
primordial density fluctuation spectrum from the galaxy survey. 
The deterministic description of bias may not be appropriate 
because of the various stochasticity of galaxy formation process. In nature, 
the biasing is epoch dependent and recent deep survey of the galaxy shows 
the large biasing at high redshift. Hence, we investigate quasi non-linear 
evolution of the stochastic bias by using the tree level perturbation method. 
Especially, the influence of the initial cross correlation on the evolution of 
the skewness and the bi-spectrum is examined in detail. 
We find that the non-linear bias can be generated dynamically. 
The small value of the initial cross correlation can bend the $\dg$-$\dm$ relation 
effectively and easily lead to the negative curvature ($b_2<0$). 
We also propose a method to predict the bias, cross correlation and 
skewness at high redshift. 
As an illustration, the possibility of the large biasing at 
high redshift 
is discussed. Provided the present bias parameter as $b=1.5$ and 
$\Omega=1.0$, we predict the large scale bias as $b=4.63$ at $z=3$ 
by fitting the bi-spectrum to the Lick catalog data. 
Our results will be important for the future deep sky survey. 
\end{abstract}
%
%
%
%%%%%%%%%%%%%%%%%%%%%%%%%%%%%%%%%%%%%%%%%%%%%%%%%%%%%%%%%%%%%%%%%%%%%%
%%%%%%%%%%%%%%%%%%%%%%%%%%%%%%%%%%%%%%%%%%%%%%%%%%%%%%%%%%%%%%%%%%%%%%
\section{Introduction}
\label{sec: intro}
%%%%%%%%%%%%%%%%%%%%%%%%%%%%%%%%%%%%%%%%%%%%%%%%%%%%%%%%%%%%%%%%%%%%%%
%%%%%%%%%%%%%%%%%%%%%%%%%%%%%%%%%%%%%%%%%%%%%%%%%%%%%%%%%%%%%%%%%%%%%%
%
%
%
%%%%%%%%%%%%%%%%%%%%%%%%
The large scale structure in the universe has evolved from the 
primordial mass fluctuations according to the gravitational instability. 
In standard picture of structure formation, the density fluctuations 
are produced by the quantum fluctuation during the inflationary 
stage of the early universe. The CMB anisotropy observed by COBE 
reflects such fluctuations at the last scattering surface. Subsequently, 
the growth of the mass fluctuation leads to formation of galaxies 
and evolution of galaxy distribution 
which is observed by the galaxy survey such as APM, 
CfA survey, {\it etc}.  
The most important issue in modern cosmology is to construct a 
consistent history of our universe from inflation to galaxy clustering. 

%%%%%%%%%%%%%%%
Statistical quantities of large scale structure provide us many important 
information to understand the evolution of our universe. 
From the power spectrum or the two-point correlation function of galaxies, 
we can evaluate the density parameter $\Omega_0$ by 
comparing the observations in the redshift space with those 
in the real space   
(\cite{H97}). In general, the $n$-point galaxy 
correlation functions characterize the large scale structure of the universe.  
These quantities play a significant role to test the prediction of 
inflation theory. 
%%%%%%%%%%%%%%%
However, there exists an uncertainty in the relation between 
the distribution of galaxy and that of total mass in the universe. 
Galaxies have been formed at the high density regions of the mass 
fluctuations. Due to the lack of our knowledge about galaxy formation 
process, we do not have the definite answer how much fraction of the total 
mass is the luminous matter. This uncertainty is quantified by the bias 
parameter of the galaxy distribution, which significantly affects 
the determination of cosmological parameters and the tests of inflation theory. 

%%%%%%%%%%%%%%%
To proceed further, it is necessary to model the galaxy distribution in 
the given mass distribution.
The linear bias gives a simple relation 
between the galaxy and mass when both fluctuations are small enough. 
Denoting the fluctuation of total mass as 
$\delta_m$ and that of the galaxy mass as $\delta_g$,
we have 
\begin{equation}
\delta_g=b~\delta_m, 
  \label{linear-bias}
\end{equation}
where $b$ is the linear bias parameter. The assumption 
(\ref{linear-bias}) yields a degeneracy of the parameters, 
$\beta(\Omega_0)=\Omega_0^{0.64}/b$,  
for determining $\Omega_0$ (\cite{H97}). We need another observation in
 order to lift up this degeneracy.  Moreover, we should keep in mind that 
the relation (\ref{linear-bias}) cannot be static. 
Indeed, the bias parameter changes in time by gravitational force 
even if the galaxy formation turned off. 
The recent observation of the high redshift galaxy survey suggests 
that the galaxies at $z\simeq3$ are largely biased, whose bias parameter 
is evaluated as $b\simeq 6$, even if we do not require the bias at present 
(\cite{P98}). 
It is necessary to consider the bias effect by taking into account 
the time evolution. 

%%%%%%%%%%%%%%%
The linear bias is appropriate as long as the linear theory 
is a good approximation to the observation. A naive extension of 
biasing to the non-linear regime is the deterministic 
non-linear bias. Suppose that the galaxy distribution $\delta_g$ 
is a local function of $\delta_m$, 
we express $\delta_g$ in powers of $\delta_m$ (\cite{FG93}) :
\begin{equation}
\delta_g=f(\delta_m)=\sum_n\frac{b_n}{n!}(\delta_m)^n.
  \label{nl-bias}
\end{equation}
The non-linear bias parameter $b_n$ appears when we compare the higher 
order statistics of the galaxy distribution with that of the mass 
fluctuation such as the skewness ($S_{3}$) and the kurtosis ($S_{4}$) 
defined by $S_{3}\equiv\langle\delta^3\rangle/\langle\delta^2\rangle^2$ 
and $S_{4}\equiv(\langle\delta^4\rangle-3\langle\delta^2\rangle^2)
/\langle\delta^2\rangle^2$, respectively. The relation becomes 
\begin{eqnarray}
&&S_{3,g}\equiv\frac{\langle\delta_g^3\rangle}{\langle\delta_g^2\rangle}^2
=b^{-1}(S_{3,m}+3\frac{b_2}{b})+{\cal O}(\langle\delta_m^2\rangle).
\label{nl-bias-skewness}
\\
&&S_{4,g}\equiv\frac{\langle\delta_g^4\rangle-3\langle\delta_g^2\rangle^2}
{\langle\delta_g^2\rangle^3}
=b^{-2}(S_{4,m}+12\frac{b_2}{b}S_{3,m}+12\frac{b_2^2}{b^2}
+4\frac{b_3}{b})+{\cal O}(\langle\delta_m^2\rangle), 
\label{nl-bias-kurtosis}
\end{eqnarray}
where we put $b_1=b$. 
Subscript $g$ and $m$ denote the statistics of the galaxies and the mass 
fluctuation, respectively. Other than the above one-point functions, 
we can calculate the amplitude of the galaxy three point correlation function 
(bi-spectrum) $Q_g$ which also has the relation $Q_g=Q_m/b+b_2/b^2$ 
(\cite{F94}). 
Since the statistical quantities for mass distribution can be obtained 
analytically under the assumption of gravitational instability, 
we can determine the bias parameters $b, b_2, b_3, \cdots,$ 
by combining the observation of $S_{g,3}, S_{g,4}$ and with that of $Q_g$ 
(\cite{FG94},\cite{GF94}). 
As for the time evolution of the bias parameter, Fry has 
discussed its influence on the skewness and the bi-spectrum 
(\cite{F96}). 
He found that $S_{3,g}$ and $Q_g$ asymptotically approach the constant 
values of $S_{3,m}$ and $Q_m$, respectively. 

%%%%%%%%%%%%%%%
There has been some confusions for the galactic bias effect 
since Kaiser provided a simple bias mechanism (\cite{K84}). 
His original idea is to explain the enhancement of the two point 
correlation function of 
the rich clusters using the statistical feature of galaxy distribution. 
Assuming the rich clusters are formed where the fluctuation $\delta_g$ 
averaged over a larger scale exceeds a certain threshold, 
he derived a simple relation between the two-point correlation function 
of galaxies and that of rich clusters. It should be stressed that 
the bias parameter introduced by Kaiser is not defined 
by density fluctuations but correlation functions. 
In our case, the galaxy correlation function may relate with the 
mass correlation function, however, 
the deterministic relation between the variables $\dg$ and $\dm$ does not 
necessarily follow from the relation between correlation functions. 

%%%%%%%%%%%%%%%
According to the inflationary scenario for the  
structure formation, the primordial density fluctuation 
is generated quantum mechanically and it is likely to have stochastic nature.
(\cite{Linde}). 
Recently, the galaxy formation process is recognized as 
non-linear and stochastic one (\cite{CO97}). Therefore we should 
treat both $\delta_g $ and $\delta_m$ as independent stochastic variables. 
The stochastic treatment of 
the bias effect is referred to as the stochastic bias, which has been 
introduced by Dekel and Lahav (\cite{Dekel}, \cite{Lahav}). 
In stochastic bias, the relation between the galaxy and the total mass 
can be described by the correlation functions. When the fluctuations 
are small enough and the linear perturbation is valid, we need three 
parameters defined below:
\begin{equation}
  \sigma^2=\langle\delta_m^2\rangle,~~~~
  b^2=\frac{\langle\delta_g^2\rangle}{\langle\delta_m^2\rangle},~~~~ 
  r=\frac{\langle\delta_m\delta_g\rangle}
  {(\langle\delta_m^2\rangle\langle\delta_g^2\rangle)^{1/2}}.
\end{equation}
The important ingredient of the stochastic bias is 
the cross correlation $r$ which is absent in the simple relation 
of linear bias (\ref{linear-bias}). 
This is an extension of the Kaiser's bias prescription. 
Matarrese {\it et al}. shows that the cross correlation between 
the regions with different thresholds of density peak naturally arises 
in general situation including both Gaussian and non-Gaussian spatial 
distribution (\cite{MLB86}). 
Mo \& White discussed the spatial distribution of galactic halloes using the 
Press \& Schechter formalism and evaluated the cross correlation by 
comparison with numerical simulation (\cite{MW96}). Recently, Catelan {\it et al}. 
extended this formalism (\cite{catelan-a}, \cite{catelan-b}, \cite{Porciani}). 
We think that the cross correlation plays an important role to describe 
the galactic bias effect. 

%%%%%%%%%%%%%%%
A naive introduction of cross correlation increases the 
unknown parameters and may complicate the determination of the cosmological 
parameters from the measurement of galaxy survey. 
Pen has discussed how to determine the parameters 
$b,~r,~\sigma$ and the other non-Gaussian variables. 
He concluded that the observation of redshift space 
distortion is useful to understand these parameters and the density 
parameter $\Omega_0$ (\cite{Pen97}). However he did not consider the 
time evolution. On the other hand, Tegmark \& Peebles 
studied the linear evolution of the stochastic bias after the galaxy 
formation epoch and on-going the galaxy formation (\cite{TP98}). 
They found that 
the bias parameter $b$ and the cross correlation $r$ approaches unity 
 even if galaxy formation never ends. 
However, their analysis is restricted to the linear evolution. 

%%%%%%%%%%%%%%%
Purpose of the present paper is to understand the 
quasi non-linear time evolution 
of the stochastic bias by observing the three-point correlation functions 
(\cite{F94}). 
The future development of 
galaxy survey such as Sloan Digital Sky Survey(SDSS) and two degree 
field survey (2dF) will provide us an enormous data of the large 
scale structure. 
In the future, 
the time evolution of galaxy distribution will be observed. Study of 
the bias evolution is of importance to separate the effect of 
the time evolution due to the galaxy formation process and the growth due to 
gravity. To investigate this, we must understand the qualitative 
and/or quantitative behavior of higher order statistics under the influence of 
gravity and recognize a role of the initial cross correlation which is 
important quantity in the stochastic bias. 

%%%%%%%%%%%%%%%
In this paper, we shall pay an attention to the 
time evolution of skewness and bi-spectrum in the stochastic 
description of the bias effect. We discuss the basic formalism of 
stochastic bias and derive the evolution equations for galaxy 
and mass distribution 
in Sec.\ref{sec: formalism}. Based on this formulation, 
we first evaluate the bias parameter and the cross correlation 
at the tree level. In Sec.\ref{sec: higher}, we develop 
the second order perturbation and study the time evolution of 
three-point function. The effect of the initial cross correlation is 
investigated by comparing with the various choices of the parameters. 
We find that the initial cross correlation significantly affects the non-linear 
relation between $\dm$ and $\dg$ which differs from the deterministic bias. 
Sec.\ref{sec: discuss} is devoted to the discussion 
about the observation from the point of the stochastic biasing. 
Using the Lick catalog data 
and the present bias parameter inferred by the {\it IRAS} galaxy survey, 
we shall show that the high redshift 
galaxies are strongly biased, which will be observed by the future 
galaxy survey. 
We summarize the results briefly and present the future prospect 
in Sec.\ref{sec: summary}. 

%
%
%
%
%%%%%%%%%%%%%%%%%%%%%%%%%%%%%%%%%%%%%%%%%%%%%%%%%%%%%%%%%%%%%%%%%%%%%%
%%%%%%%%%%%%%%%%%%%%%%%%%%%%%%%%%%%%%%%%%%%%%%%%%%%%%%%%%%%%%%%%%%%%%%
\section{Basic Formulation of Stochastic Bias}
\label{sec: formalism}
%%%%%%%%%%%%%%%%%%%%%%%%%%%%%%%%%%%%%%%%%%%%%%%%%%%%%%%%%%%%%%%%%%%%%%
%%%%%%%%%%%%%%%%%%%%%%%%%%%%%%%%%%%%%%%%%%%%%%%%%%%%%%%%%%%%%%%%%%%%%%
%
%
%
This section is devoted to a basic formulation of stochastic bias 
for further analysis of next section. 
We first consider the stochastic description of fluctuation for the galaxy 
distribution and the total mass distribution. In the case of weakly non-linear 
evolution, 
the generating functional for the stochastic variables of galaxy and 
total mass is constructed. Subsequently, we give the 
evolution equations and initial conditions. We will analyze these equations 
perturbatively and discuss the linear evolution of the stochastic bias. 
%
%
%
%%%%%%%%%%%%%%%%%%%%%%%%%%%%%%%%%%%%%%%%%%%%%%%%%%%%%%%%%%%%%%%%%%%%%%
\subsection{Stochastic description}
%%%%%%%%%%%%%%%%%%%%%%%%%%%%%%%%%%%%%%%%%%%%%%%%%%%%%%%%%%%%%%%%%%%%%%
%
%
%
%
%%%%%%%%%%%%%%%
The basic quantities to treat the stochastic bias are the fluctuations 
of total mass and the 
galaxy distribution, $\delta_m$ and $\delta_g$. They are 
defined by 
\begin{equation}
  \dm(x)\equiv \frac{\rho(x)-\bar{\rho}}{\bar{\rho}}, ~~~~~
  \dg(x)\equiv \frac{n_g(x)-\bar{n}_g}{\bar{n}_g},
\end{equation}
where barred quantities means homogeneous averaged density. 
We shall investigate the skewness and the bi-spectrum of the 
galaxy distribution $\dg$. We evaluate them using the smoothed 
density field by the window function:
\begin{equation}
  \delta_{m,g}(R)\equiv\int d^3x \cdot W_R(x)\delta_{m,g}(x).
\end{equation}
The window function $W_R(x)$ we adopt here is the spherical top-hat 
smoothing with the radius $R$ defined bellow:
\begin{eqnarray}
&&  W_R(x)=\left\{
\begin{array}{c}
1~~~~~(|x|\leq R), \\
0~~~~~(|x|>  R) .
\end{array}
\right.
\end{eqnarray}
%

%%%%%%%%%%%%%%%
As we are interested in the time evolution of bias effect 
under the influence of gravity, we ignore the galaxy formation process 
in consideration of the evolution of galaxy distribution. 
The stochastic bias is described by the stochastic variables 
 $\dm$ and $\dg$. Their stochasticity is 
determined by the probability distribution functional 
${\cal P}[\dm(x),~\dg(x)]$. Once ${\cal P}[\dm(x),~\dg(x)]$ is given, 
all the information of the distribution functional is encoded in 
the correlation functions of $\dm$ and $\dg$,  
and all of the correlation functions can be obtained from 
the following generating functional: 
\begin{eqnarray}
 && {\cal Z}[J_m,J_g]=\int\int{\cal D}\delta_m(x){\cal D}\delta_g(x)
\cdot{\cal P}[\delta_m(x),\delta_g(x)]
\nonumber \\
&&~~~~~~~~~~~~~~~~~~~~~~~~~\times
\exp{\left[i\int d^3x W_R(x)
    \left\{J_m\delta_m(x)+J_g\delta_g(x)\right\}\right]},
\label{Z-1}
\end{eqnarray}
where $J_m,~J_g$ are external source terms. 
Using (\ref{Z-1}), the $n$-th moment of the filtered one-point function 
is written by 
\begin{eqnarray}
&&  \langle\{\delta_m(R)\}^j\cdot\{\delta_g(R)\}^{n-j}\rangle
=i^{-n}\left.
\frac{\delta^n {\cal Z}}{\delta{J}_m^j\cdot\delta{J}_g^{n-j} }
\right|_{J_m=J_g=0}.
\end{eqnarray}

%
%
%%%%%%%%%%%%%%%%%%%%%%%
The probability functional ${\cal P}[\dm(x),\dg(x)]$ can be determined 
in principle by the study of galaxy formation and evolution of 
fluctuations before galaxy formation. 
In our prescription of stochastic bias, we give ${\cal P}[\dm(x),\dg(x)]$ 
as a parameterized function. A simple choice is Gaussian:
\begin{equation}
  {\cal P}[\delta_m(x),\delta_g(x)]={\cal N}^{-1}
\exp{\left[-(\delta_m,\delta_g)\mbox{\boldmath$G$}^{-1}
  \left(
\begin{array}{c}
\delta_m \\ \delta_g
\end{array}
\right)
\right]},
\end{equation}
where $\mbox{\boldmath$G$}$ is a $2\times2$ matrix and ${\cal N}$ is 
a normalization constant. The non-vanishing off-diagonal component 
of $\mbox{\boldmath$G$}$ represents the cross correlation 
between $\delta_m$ and $\delta_g$. We obtain
\begin{equation}
  \langle\dm(R)\dg(R)\rangle=
  \int d^3x d^3y \cdot W_R(x)W_R(y)\{\mbox{\boldmath$G$}\}_{12}
    \dm(x)\dg(x).
\end{equation}
To take into account the non-linearity of gravitational evolution and 
the formation process of galaxies, the Gaussian 
statistics does not lead to a correct description. 
Pen has discussed the deviation from the Gaussian distribution 
by using Edgeworth expansion(\cite{JWACB95}),  
although his analysis is restricted to the static case (\cite{Pen97}). 

%%%%%%%%%%%%%%%%%%%%%%%
Since the density fluctuations are small on large scales, 
perturbative analysis is applicable to study the evolution of density 
fluctuation. 
In this case, the non-Gaussian 
stochastic distribution for $\dm$ and $\dg$ can be expressed in powers of 
the Gaussian variables $\Dm $ and $\Dg$ whose variances are 
small enough. We shall write the initial condition at 
the end of galaxy formation ($t=t_i$) as
\begin{equation}
  \delta_m=f(\Delta_m),~~~~~\delta_g=g(\Delta_g).   
\label{initial-con}
\end{equation}
If we know the function $f$ and $g$ and stochastic property for 
$\Dm$ and $\Dg$, the initial condition for $\dm$ and $\dg$ 
is determined. 
The time evolution may change the initial statistics. We can obtain 
the correlation functions after the galaxy formation epoch 
by solving the evolution equations for $\dm$ and $\dg$. 
Then the time-dependent generating functional becomes 
\begin{eqnarray}
&&{\cal Z}[J_m,J_g;t]=\int\int{\cal D}\Delta_m(x){\cal D}\Delta_g(x)
\cdot
{\cal N}^{-1}\exp\left[-(\Delta_m,\Delta_g)\mbox{\boldmath$\tilde{G}$}^{-1}
\left(
\begin{array}{c}
\Delta_m \\ \Delta_g
\end{array}
\right)\right]
\nonumber 
\\
&&~~~~~~~~~~~~~~\times
\exp{\left[i\int d^3x W_R(x)\left\{J_m\delta_m(\Dm,\Dg,t)+J_g
      \delta_g(\Dm,\Dg,t)\right\}\right]}.
\label{Z-2}
\end{eqnarray}
The time dependence of $\dm$ and $\dg$ are obtained from 
the perturbative expansion:
\begin{equation}
  \delta_m(\Delta_{m},\Dg,t)=\delta_m^{(1)}+\delta_m^{(2)}+\cdots,~~~
  \delta_g(\Delta_m,\Dg,t)=\delta_g^{(1)}+\delta_g^{(2)}+\cdots. 
\label{expansion}
\end{equation}
Since the $n$-th order perturbed variables $\delta_{m,g}^{(n)}$ 
have the same order of magnitude as $(\Delta_{m,g})^n$,  
the initial condition given by (\ref{initial-con})  
assigns the solutions (\ref{expansion}) order by order. 
From (\ref{Z-2}) and (\ref{expansion}), the lowest order non-vanishing 
$n$-th moment can be evaluated from the results up to the 
$(n-1)$-th order perturbation. 
It is composed of ``tree'' diagram with no loop, in 
graphical representation of the perturbation theory (\cite{F84}). 
The tree-level third order moment for galaxy distribution becomes 
\begin{equation}
  \langle[\delta_g(R)]^3\rangle=
i^{-3}\left.\frac{\delta^3{\cal Z}}{\delta J_g^3}\right|_{J_g=J_m=0}
\simeq
3\langle[\delta_g^{(1)}(R)]^2\delta_g^{(2)}(R)\rangle.
\label{third-moment}
\end{equation}
From (\ref{Z-2}), in principle, we can obtain the joint distribution 
function of $\dm$ and $\dg$ by Laplace transformation.  
%
%
%
%%%%%%%%%%%%%%%%%%%%%%%%%%%%%%%%%%%%%%%%%%%%%%%%%%%%%%%%%%%%%%%%%%%%%%
\subsection{Evolution equations and initial condition}
%%%%%%%%%%%%%%%%%%%%%%%%%%%%%%%%%%%%%%%%%%%%%%%%%%%%%%%%%%%%%%%%%%%%%%
%
%
%
%%%%%%%%%%%%%%%%%%%%%%%%%%%%%
The evolution of the galaxy and total mass distribution after 
galaxy formation is determined by gravity including the effect of 
cosmic expansion. The total mass density $\rho(x)$ is approximated by 
a non-relativistic pressureless fluid. The averaged homogeneous part 
$\bar{\rho}$ is obtained by solving the FRW equations:
\begin{eqnarray}
&&  \left(\frac{\dot{a}}{a}\right)^2\equiv H^2=
  \frac{8\pi G}{3}\bar{\rho}+\frac{\Lambda}{3}-\frac{K}{a^2}~~;
~~(K=0,\pm1),  
\\
&&~~\frac{\ddot{a}}{a}=-\frac{4\pi G}{3}\bar{\rho}+\frac{\Lambda}{3},  
\end{eqnarray}
where $a$ is the expansion factor, $K$ is the spatial curvature, 
and $\Lambda$ is a cosmological constant. 
Using these variables, the density parameter $\Omega$ is defined by 
\begin{equation}
\Omega\equiv \frac{8\pi G}{3}\frac{\bar{\rho}}{H^2}.  
\end{equation}

%
%%%%%%%%%%%%%%%%%%%%%%%%%%%%%
Let us consider the fluctuating part of the mass distribution $\dm$. 
It obeys the equation of continuity 
and the peculiar velocity field $\v$ is determined by the Euler equation 
in the presence of gravitational potential. On large scales, 
the assumption that the velocity field is irrotational would be valid. 
Then, the basic quantities for describing the dynamics are 
reduced to $\dm$ and 
$\theta\equiv\nabla\cdot\v/(aH)$. The evolution equations 
become (\cite{Peebles})
\begin{eqnarray}
&& \frac{\partial\delta_m}{\partial t}+H\theta+
\frac{1}{a}\nabla\cdot(\delta_m\v)=0,
\label{basic-eq1}
\\
&&  \frac{\partial\theta}{\partial t}+\left(1-\frac{\Omega}{2}+
\frac{\Lambda}{3H^2}\right)H\theta+\frac{3}{2}H\Omega\delta_m
+\frac{1}{a^2H}\nabla\cdot(\v\cdot\nabla)\v=0.
\label{basic-eq2}
\end{eqnarray}

%
%%%%%%%%%%%%%%%%%%%%%%%%%%%%%
As for the galaxy distribution, $\dg$ should satisfy equation of 
continuity as long as the galaxy formation is not efficient. 
After formation epoch, the galaxies move along the velocity field 
determined by the gravitational potential. Therefore we have (\cite{F96})  
\begin{equation}
  \frac{\partial\delta_g}{\partial t}+H\theta+
\frac{1}{a}\nabla\cdot(\delta_g\v)=0.
\label{basic-eq-gal}
\end{equation}
Eqs.(\ref{basic-eq1}), (\ref{basic-eq2}), and (\ref{basic-eq-gal}) 
are our basic equations to determine the evolution of $\dm$ and $\dg$. 
%%%%%%%%%%%%%%%%%%%%%%%%%%%%%

We next consider the initial condition (\ref{initial-con}) 
given at the end of galaxy formation $t=t_i$. 
For the total mass fluctuation, we believe that 
$\dm$ is produced during the very early stage of the universe. In 
standard scenario of the inflationary universe, the density fluctuation 
is generated by the quantum fluctuation and may have the Gaussian statistics. 
We regard such fluctuations as $\Dm$. After the inflation, 
the gravitational instability incorporates the deviation from the Gaussian 
fluctuation. Since the galaxy formation does not affect the 
evolution of $\dm$ on large scales, the total mass fluctuation $\dm$ at 
the end of galaxy formation can be determined by the gravitational 
instability. At that time, the growing mode is dominant. Therefore,   
we obtain the initial condition $\dm=f(\Dm)$ from the evolution 
equations by dropping the decaying mode. 

%%%%%%%%%%%%%%%%%%%%%%%%%%%%%
On the other hand, the fluctuation of galaxy number density is induced 
by the galaxy formation. To determine the function $g(\Dg)$, we need 
to know halo formation processes. An analytic model for the spatial 
clustering of haloes is discussed by Mo \& White and several authors 
(\cite{MW96}). 
Here, we rather treat $g(\Dg)$ as a parameterized function, whose 
unknown parameters are determined by the observation of galaxy survey. 
Assuming $g(\Dg)$ as a local function of $\Dg$, we have 
\begin{equation}
  g(\Dg)=\Dg+\frac{h}{6}(\Dg^2-\langle\Dg^2\rangle)+\cdots.
\label{init-g}
\end{equation}

%
%%%%%%%%%%%%%%%%%%%%%%%%%%%%%
The remaining task is to specify  
the stochastic property of $\Dm$ and $\Dg$. Since these are the Gaussian 
variables given by (\ref{Z-2}), their statistics can be characterized completely 
by the three parameters below: 
\begin{equation}
  \sigma_0^2={\langle\Delta_m^2\rangle},~~~~
  b_0^2=\frac{\langle\Delta_g^2\rangle}{\langle\Delta_m^2\rangle},~~~~
  r_0=\frac{\langle\Delta_g\Delta_m\rangle}
{\left(\langle\Delta_m^2\rangle\langle\Delta_g^2\rangle\right)^{1/2}},
\label{stochastic-para}
\end{equation}      
which is equivalent to determining the matrix 
$\tilde{\mbox{\boldmath$G$}}$. Notice that $-1\leq r_0\leq1$ can be deduced 
from the Schwarz inequality. We simply assume that the parameters 
$b_0,~r_0$ are constant, which comes from the fact that there is no 
evidence of the scale-dependent bias on large scales (\cite{Mann97}). 
Therefore 
power spectra for the galaxy two-point correlation function 
and the galaxy-mass cross correlation are simply represented by the power 
spectrum $P(k)$ of the total mass fluctuation given by 
\begin{equation}
  \sigma_0^2(R)=\int\frac{d^3\k}{(2\pi)^3}\tilde{W}_R^2(kR)P(k), 
\label{P-and-sigma}
\end{equation}
where $\tilde{W}_R(kR)$ is the Fourier transform of the top-hat window 
function,
\begin{equation}
\tilde{W}_R(kR)=\frac{2}{(kR)^3}\left[\sin{(kR)}-kR\cos{(kR)}\right].
\end{equation}
%
%
%
%
%
%%%%%%%%%%%%%%%%%%%%%%%%%%%%%%%%%%%%%%%%%%%%%%%%%%%%%%%%%%%%%%%%%%%%%%
\subsection{Variance and covariance} 
\label{subsec: Linear}
%%%%%%%%%%%%%%%%%%%%%%%%%%%%%%%%%%%%%%%%%%%%%%%%%%%%%%%%%%%%%%%%%%%%%%
%
%
%
%%%%%%%%%%%%%%%%%%%%%%%%%%%%%
We are in a position to study the evolution of stochastic bias. 
To begin with, it is necessary to study the variance and covariance. 
To calculate these quantity at the tree level, the linear perturbation 
is sufficient. 
In next section, we shall develop the second order perturbation and 
analyze the skewness and bi-spectrum. 

%%%%%%%%%%%%%%%%%%%%%%%%%%%%%
Initial conditions and the evolution equations 
(\ref{basic-eq1}), (\ref{basic-eq2}), and (\ref{basic-eq-gal}) yield 
the linear order solutions:
\begin{eqnarray}
\delta_m^{(1)}(x,t)&=&\Delta_m(x) D_+(t),
\label{m-sol-1st}
 \\
\delta_g^{(1)}(x,t)&=&\Delta_m(x) (D_+(t)-1)+\Delta_g(x),
\label{g-sol-1st}
\end{eqnarray}
where the function $D_+(t)$ denotes the solution of growing mode 
by setting $D_+(t_i)=1$, which satisfies  
\begin{equation}
  \ddot{D}_++2H\dot{D}_+-\frac{3}{2}H^2\Omega D_+=0. 
\label{linear-pertur}
\end{equation}
In Einstein-de Sitter universe ($\Omega=1$), 
we have $D_+(t)=a(t)/a(t_i)$. 

%%%%%%%%%%%%%%%%%%%%%
Since the stochastic property is already given by (\ref{Z-2}) and
(\ref{stochastic-para}), we can obtain the relation between 
$\dm$ and $\dg$ after galaxy formation. 
Similar to the expression (\ref{stochastic-para}), 
we define the time-dependent parameters as follows: 
\begin{equation}
 \sigma^2(t)\equiv{\langle\delta_m^2(t)\rangle},~~~~
b^2(t)\equiv\frac{\langle\delta_g^2(t)\rangle}
{\langle\delta_m^2(t)\rangle},~~~~
r(t)\equiv\frac{\langle\delta_g(t)\delta_m(t)\rangle}
{\left(\langle\delta_m^2(t)\rangle\langle\delta_g^2(t)\rangle\right)^{1/2}}
\label{t-dependent-stochastic-para}
\end{equation}      
Substituting (\ref{m-sol-1st}) and (\ref{g-sol-1st}) into 
(\ref{t-dependent-stochastic-para}), we get 
\begin{eqnarray}
\sigma(t)&=&\sigma_0 D_+,
  \nonumber
\\
b(t)&=&\frac{\sqrt{(D_+-1)^2+2b_0r_0(D_+-1)+b_0^2}}{D_+}, 
  \label{linear-t-bias-parameter}
\\
r(t)&=&b^{-1}(t)\left(\frac{D_+-1+b_0r_0}{D_+}\right) .
  \nonumber
\end{eqnarray}
%
%%%%%%%%%%%%%%%%%%%%%

These equations coincides with the result of Tegmark \& Peebles 
in the absence of galaxy formation (\cite{TP98}). The time dependent 
correlation functions (\ref{linear-t-bias-parameter}) show that 
the bias parameter $b(t)$ and the cross correlation $r(t)$ 
approach unity asymptotically as the fluctuations grow. 
Note that the deterministic bias relation 
restricts on the parameter 
$r_0=1$, which leads to $r(t)=1$. 
The value $r_0=0$ means that $\dm$ and $\dg$ have no correlation initially. 
Subsequent evolution of $\dg$ and $\dm$ develops the correlations and 
ultimately makes the complete correlation, i.e, $r=1$ in the case of 
Einstein-de Sitter universe. 
This indicates that the cross correlation significantly affects 
the evolution of the higher order statistics. 
The functional form of the bias is non-linear, in general. 
This relation gives the linear bias relation (\ref{linear-bias}) 
if we restrict us to the linear 
region, however, this is not sufficient obviously.  
%
%
%
%
%%%%%%%%%%%%%%%%%%%%%%%%%%%%%%%%%%%%%%%%%%%%%%%%%%%%%%%%%%%%%%%%%%%%%%
%%%%%%%%%%%%%%%%%%%%%%%%%%%%%%%%%%%%%%%%%%%%%%%%%%%%%%%%%%%%%%%%%%%%%%
\section{Skewness and Bi-spectrum}
\label{sec: higher}
%%%%%%%%%%%%%%%%%%%%%%%%%%%%%%%%%%%%%%%%%%%%%%%%%%%%%%%%%%%%%%%%%%%%%%
%%%%%%%%%%%%%%%%%%%%%%%%%%%%%%%%%%%%%%%%%%%%%%%%%%%%%%%%%%%%%%%%%%%%%%
%
%
Following the formulation of the stochastic bias in previous section, 
we investigate the bias effect on the skewness and the bi-spectrum. 
First of all, 
the solutions of second order perturbation are given.   
Using this result and the analysis in Sec.\ref{subsec: Linear}, 
we evaluate the time evolution of the skewness and 
bi-spectrum of galaxies and analyze the influence of cross correlation 
$r$ on them in Sec.\ref{subsec: skewness} and 
Sec.\ref{subsec: bi-spectrum}.  
To evaluate the skewness with the top-hat smoothing, 
the integration including 
the window function $W_R$ is tedious and complicated. However, the 
calculation is tractable in the Fourier space because the useful 
formulae have already been found (\cite{B94a}, \cite{B94b}). 
Here, utilizing these formulae, we develop the second order perturbation 
in the Fourier space. The variables $\dm, \dg$ and $\theta$ are expanded 
as follows:
\begin{equation}
  \delta_{m,g}(\x,t)=\int\frac{d^3\k}{(2\pi)^3}\hat{\delta}_{m,g}
  (\k,t)e^{-i \k\x},~~~
  \theta(\x,t)=\int\frac{d^3\k}{(2\pi)^3}\hat{\theta}(\k,t)e^{-i\k\x}.
\end{equation}
Hereafter, we denote the Fourier coefficient by attaching the hat. 

The calculation in the Fourier space is thoroughly 
investigated by Bernardeau, although he did not consider the 
evolution of $\dg$(\cite{B94b}). We display the results only. The 
second order solutions satisfying the initial conditions are
\begin{eqnarray}
  \hat{\delta}_m^{(2)}(\k,t)&=&\int\frac{d^3\k'}{(2\pi)^3}
  \left[D_+^2(t)\left(\frac{6}{7}{\cal P}(\k',\k-\k')
      +\frac{1}{7}{\cal P}(\k-\k',\k')-\frac{3}{2}{\cal L}(\k',\k-\k')\right)
\right.
\nonumber\\
&&~~~~~~~~~~~~~~~~~~~~~~~~~~~~~~
\left.+\frac{3}{4}E_+(t){\cal L}(\k',\k-\k')\right]\hat{\Delta}_m(\k-\k')
\hat{\Delta}_m(\k'),
\label{2nd-sol-m}
\end{eqnarray}
\begin{eqnarray}
\hat{\delta}^{(2)}_g(\k,t)&=&\hat{\delta}_m^{(2)}(\k,t)-
\hat{\delta}_m^{(2)}(\k,t_i)
\nonumber\\
&+&(D_+(t)-1)\int\frac{d^3\k'}{(2\pi)^3}{\cal P}(\k',\k-\k')
(\hat{\Delta}_m(\k')\hat{\Delta}_g(\k-\k')
-\hat{\Delta}_m(\k')\hat{\Delta}_m(\k-\k')) 
\nonumber\\
&+&\frac{h}{6}\cdot\int\frac{d^3\k'}{(2\pi)^3}
\left(\hat{\Delta}_g(\k')\hat{\Delta}_g(\k-\k')
  -\langle\hat{\Delta}_g^2\rangle\delta_D(\k')\right), 
\label{2nd-sol-g}
\end{eqnarray}
where
\begin{equation}
  {\cal P}(\k_1,\k_2)=1+\frac{(\k_1\cdot \k_2)}{|\k_1|^2},~~~~~
  {\cal L}(\k_1,\k_2)=1-\frac{(\k_1\cdot \k_2)^2}{|\k_1|^2|\k_2|^2}. 
\end{equation}
The solutions (\ref{2nd-sol-m}) and (\ref{2nd-sol-g}) contain the function 
$E_+(t)$ which satisfies $E_+(t_i)=1$. This is the inhomogeneous 
solution of the following equation:
\begin{eqnarray}
&&  \ddot{E}_++2H\dot{E}_+-\frac{3}{2}H^2\Omega E_+=
3H^2\Omega D_+^2+\frac{8}{3}\dot{D}_+^2.    
\end{eqnarray}
In Einstein-de Sitter universe, we have  
\begin{eqnarray}
E_+(t)=\frac{34}{21}D_+^2(t). 
\label{sol-E_+}
\end{eqnarray}
It is known that the $\Omega$ and $\Lambda$ dependences of $E_+/D_+^2$ 
is extremely weak (\cite{B94b}). 
Therefore, we proceed to analyze the skewness and 
bi-spectrum by replacing $E_+(t)$ with $(34/21)D_+^2(t)$.
%
%
%
%%%%%%%%%%%%%%%%%%%%%%%%%%%%%%%%%%%%%%%%%%%%%%%%%%%%%%%%%%%%%%%%%%%%%%
\subsection{Skewness}
\label{subsec: skewness}
%%%%%%%%%%%%%%%%%%%%%%%%%%%%%%%%%%%%%%%%%%%%%%%%%%%%%%%%%%%%%%%%%%%%%%
%
%
%
The second order solutions (\ref{2nd-sol-m}) and (\ref{2nd-sol-g}) 
together with the linear order solutions 
give the tree-level third order moment by substituting them into 
(\ref{third-moment}). From (\ref{Z-2}), the result of the 
smoothed one-point function becomes
\begin{eqnarray}
&&  \frac{\langle\delta_g^3(R)\rangle}{\sigma^4_0}\simeq 
\left(\frac{34}{7}-\gamma\right)(D_+^2-1)(D_+-1+b_0r_0)^2
\nonumber\\
&&~~~~+(6-\gamma)\left\{b_0r_0(D_+-1)+(D_+-1)^2\right\}
\left\{(b_0r_0-1)D_++(b_0^2+1-2b_0r_0)\right\}
\nonumber\\
&&~~~~+h\cdot I_W \left\{b_0r_0(D_+-1)+b_0^2\right\}^2,
\label{3-moment}
\end{eqnarray}
where
\begin{eqnarray}
&& \gamma=-\frac{d}{d(\log{R})}\cdot [\log{\sigma^2_0(R)}],
\label{gamma}
\\
&& I_W=\sigma_0^{-4}\int\frac{d^3\k_1d^3\k_2}{(2\pi)^6}
\tilde{W}_R(|\k_1+\k_2|R)\tilde{W}_R(k_1R)\tilde{W}(k_2R)P(k_1)P(k_2).
\label{I_W}
\end{eqnarray}
The skewness of the galaxy distribution is obtained as follows:
\begin{equation}
S_{3,g}=\frac{\langle\delta_g^3(R)\rangle}{\langle\delta_g^2(R)\rangle^2}~~;
~~~~~~\frac{\langle\delta_g^2(R)\rangle}{\sigma_0^2}
=(D_+-1)^2+2b_0r_0(D_+-1)+b_0^2.
\label{skewness-g}
\end{equation}
We see from (\ref{skewness-g}) that $S_{3,g}$ evolves from 
the initial value $h\cdot I_W$ and asymptotically approaches the skewness 
of the total mass $S_{3,m}$, 
\begin{equation}
  \label{S_m}
  S_{3,m}=\frac{34}{7}-\gamma, 
\end{equation}
which is constant in time. The numerical value $I_W$ 
comes from the contribution of non-Gaussian initial distribution. 
It is almost equal to unity for the power spectrum $P(k)\propto k^n~ 
(-3\leq n \leq -1)$, which has been checked by the Monte Carlo 
integration of (\ref{I_W}). 
%
%
%
%%%%%%%%%%%%%%%%%%%%%%%%%%

We evaluate the time evolution of skewness by plotting 
(\ref{skewness-g}) for the various parameters. 
Hereafter we assume that the power spectrum simply 
obeys the single power-law $P(k)\propto k^n$, which yields 
$\gamma=n+3$ from (\ref{gamma}). 
%%%%%%%%%%%%%%%%%%%%%%%%%%
 In Fig.1, we have plotted the evolution of $S_{3,g}$ as a function of 
the expansion factor $a$. For a specific choice 
of the parameters, we take $\Omega_0=1,~n=-3,~h=3.0$, for convinience. 
We have verified that the other choice of the parameters leads to 
the similar behavior of the time evolution of the skewness. 
The initial conditions at $a=1$ for each lines are 
$b_0=2.0, r_0=1.0$({\it solid line}), 
$b_0=2.0, r_0=0.5$({\it long-dashed line}), 
$b_0=1.0, r_0=0.0$({\it short-dashed line}) and 
$b_0=0.5, r_0=0.0$({\it dotted line}). 
%%%%%%%%%%%%%%%%%%%%%%%%%%
The figure manifests the deviation from the 
deterministic bias $r=1$ (solid and long-dashed lines). 
The skewness with the smaller 
$r_0$ tends to become the larger value. 
We see that the gravitational instability changes 
the cross correlation largely and this enhances 
the non-linear growth of skewness $S_{3,g}$. 
The smaller bias parameter $b_0$ also leads to 
the larger skewness (short-dashed and dotted lines). Especially, 
the skewness in anti-biasing case ($b_0<1$) has a characteristic 
initial behavior, since the rare distribution 
of galaxies rapidly concentrates on that of total mass due to the 
attractive force of gravity. 
%%%%%%%%%%%%%%%%%%%%%%%%%%

For more various initial data,  the resulting parameters which 
evolve from the high redshift are exhibited in Table.1. 
Provided that the initial parameters $b_0,~r_0$ and $h(=S_{3,g})$ at $z=3$,  
we obtain the bias parameter $b$ and the skewness $S_{3,g}$ at present, 
which qualitatively have the same behavior as depicted in Fig.1. 
Although we obtain the skewness $S_{3,g}$ in the treatment of the 
stochastic bias, the non-linear bias 
parameter $b_2$ which gives the deterministic $\dg$-$\dm$ relation 
\begin{equation}
  \dg=f(\dm)=
b\dm+\frac{b_2}{2}(\dm^2-\langle\dm^2\rangle)+\cdots
\end{equation}
can be evaluated by using the result of the deterministic 
bias (\ref{nl-bias-skewness}). The $\dg$-$\dm$ relation 
has to do with the peak of the joint probability distribution for 
$\dm$ and $\dg$. 
The non-vanishing value $b_2$ represents the deviation from a simple 
linear relation $\dg\propto\dm$. Table.1 says that 
the smaller cross correlation and the smaller bias parameter 
bend the $\dg$-$\dm$ relation efficiently. 
The remarkable property of the stochastic bias is that 
$b_2$ could become negative for the small initial cross correlation, 
 although it approaches zero asymptotically. 
This tendency can be interpreted in the following way. Assume 
the existence of the primary linear bias $b$, it will approach unity 
asymptotically. However, this relaxation may not proceed monotonously. 
The larger fluctuation accelerates the relaxation due to 
the non-linearity of the gravitational force. Moreover, 
the small initial cross correlation enhances this acceleration. Hence, 
$\dg$-$\dm$ relation with the negative $b_2$ can be obtained.   
Fry and Tegmark \& Peebles found that both $b$ and $r$ approach 
to unity asymptotically (\cite{F96},\cite{TP98}). Here, we found 
the general tendency of the  non-linear bias which is determined 
by the gravitational dynamics. 
The tilted relation with a negative $b_2$ is consistent with the recent 
numerical simulation (\cite{DL98}). 
%%%%%%%%%%%%%%%%%%%%%%%%%%

Table.1 also gives the cases for the different power spectrum from 
the scale invariant one ($n=-3$) and the different density parameter from 
$\Omega_0=1$. 
From the evolution of skewness with the spectral index $n=-1$, 
we observe that the initial non-linear bias parameter $b_2$ 
rapidly decreases to zero compared with the case with index $n=-3$. 
Since the spectrum with $n=-1$ has larger power than $n=-3$ on 
large scales, the result shows that the tilted spectrum could enhance 
the non-linear relation of $\dg$-$\dm$. 
As for the low density universe($\Omega_0=0.2$), the 
growth of the density fluctuations becomes slow and the variations of 
$b$ and $b_2$ are small. Accordingly,  
the deviations from $b=1$ and $b_2=0$ in low density universe are 
larger than that in Einstein-de Sitter universe.
%
%
%
%
%
%
%%%%%%%%%%%%%%%%%%%%%%%%%%%%%%%%%%%%%%%%%%%%%%%%%%%%%%%%%%%%%%%%%%%%%%
\subsection{Bi-spectrum}
\label{subsec: bi-spectrum}
%%%%%%%%%%%%%%%%%%%%%%%%%%%%%%%%%%%%%%%%%%%%%%%%%%%%%%%%%%%%%%%%%%%%%%
%
%
In addition to the skewness, the bi-spectrum provides us the other 
information of three-point correlation function. In Fourier space, 
it is defined by 
\begin{eqnarray}
&& \langle\delta_g(\k_1)\delta_g(\k_2)\delta_g(\k_3)\rangle
=\delta_D(\k_1+\k_2+\k_3)B_g(\k_1,\k_2,\k_3).  
\end{eqnarray}
The amplitude $B_g$ relates with the third moments (\ref{3-moment}) 
as follows:
\begin{equation}
  \langle\delta_g^3(R_0)\rangle
  =\int\frac{d^3\k_1}{(2\pi)^3}\int\frac{d^3\k_2}{(2\pi)^3}
\tilde{W}(|\k_1+\k_2|R_0)\tilde{W}(k_1R_0)\tilde{W}(k_2R_0)
B_g(\k_1,\k_2,\k_3). 
\end{equation}
Recall that $\k_1+\k_2+\k_3=0$, we obtain 
\begin{eqnarray}
  \label{B_g}
&&  B_g(\k_1,\k_2,\k_3)=\left[
\{{\cal P}(\k_1,\k_2)+{\cal P}(\k_2,\k_1)
-\frac{4}{7}{\cal L}(\k_1,\k_2)\}  (D_+^2-1)(D_+-1+b_0r_0)^2\right.
\nonumber
\\
&&~~~~~~+\left.\{{\cal P}(\k_1,\k_2)+{\cal P}(\k_2,\k_1)\}
(D_+-1+b_0r_0)\{(b_0r_0-1)D_++b_0^2+1-2b_0r_0\}\right.
\nonumber
\\
&&~~~~~~+\left.\frac{h}{3}(b_0r_0(D_+-1)+b_0^2)^2\right]P(k_1)P(k_2)
\nonumber
\\
&&~~~~~~+~~\mbox{permutations}. 
\label{B-amplitude}
\end{eqnarray}
$P(k)$ is the power spectrum of total mass fluctuations given by 
(\ref{P-and-sigma}). A more convenient observable is the reduced 
bi-spectrum amplitude $Q_g$ which is briefly discussed in Sec.
\ref{sec: intro} (\cite{F94}, \cite{F96}):
\begin{equation}
Q_g\equiv
\frac{B_g(\k_1,\k_2,\k_3)}
{P_g(k_1)P_g(k_2)+P_g(k_2)P_g(k_3)+P_g(k_3)P_g(k_1)}. 
\label{Q_g}
\end{equation}
The power spectrum of the galaxies $P_g(k)$ is written by 
\begin{equation}
  P_g(k)=\left\{(D_+-1)^2+2b_0r_0(D_+-1)+b_0^2\right\}^2 P(k).
\end{equation}

The reduced bi-spectrum amplitude $Q_m$ for the mass fluctuation 
has a remarkable property that it is scale independent and constant 
in time. In our case, $Q_g$ is also scale independent due to 
the assumption (\ref{stochastic-para}), although it depends on time and 
approaches to $Q_m$. To evaluate the configuration shape of 
the bi-spectrum, we plot $Q_g$ as a function of the angle 
$\theta=\cos^{-1}(\hat{\k}_1\cdot\hat{\k}_2)$ by setting $k_1/k_2=2$.  
Fig.2 shows the various cases of the present amplitude $Q_g~~(z=0)$ whose 
initial conditions are set at $z=3$  in the case of 
Einstein-de Sitter universe($\Omega_0=1$). We also take the initial skewness 
parameter $h=3.0$ and the spectral index $n$ is chosen as $n=-3$. 
In Fig.2a, we plot the bi-spectrum for the various initial values of the 
cross correlation when the initial bias parameter is fixed ($b_0=2.0$). 
Each solid line shows the result of the initial conditions 
$r_0=0.0,~0.25,~0.5,~0.75$, and $1.0$({\it from up to bottom}), respectively. 
We see that the cross correlation affects the shape of the bi-spectrum. 
For the smaller $r_0$, the shape of $Q_g$ becomes steeper and it approaches 
$Q_m$ faster. This is the same behavior as the non-linear growth of 
$S_{3,g}$. In cases the initial cross correlation is fixed, 
the shape of the bi-spectrum 
changes significantly depending on the bias parameters 
(Fig.2b). In the next section, we will show that the small shape 
dependence of the cross correlation plays an important role for 
explaining both the present observation of galaxies and the galaxies 
at high redshift. 
%
%
%
%
%
%%%%%%%%%%%%%%%%%%%%%%%%%%%%%%%%%%%%%%%%%%%%%%%%%%%%%%%%%%%%%%%%%%%%%%
%%%%%%%%%%%%%%%%%%%%%%%%%%%%%%%%%%%%%%%%%%%%%%%%%%%%%%%%%%%%%%%%%%%%%%
\section{Discussion}
\label{sec: discuss}
%%%%%%%%%%%%%%%%%%%%%%%%%%%%%%%%%%%%%%%%%%%%%%%%%%%%%%%%%%%%%%%%%%%%%%
%%%%%%%%%%%%%%%%%%%%%%%%%%%%%%%%%%%%%%%%%%%%%%%%%%%%%%%%%%%%%%%%%%%%%%
%
%
%
In previous section, we studied the influence of the cross correlation 
on the time evolution of skewness and bi-spectrum. In this section, 
using these results, we describe how the stochastic treatment 
of the bias affects the analysis of observational data for 
the galaxy distribution. We will present a method for predicting the 
bias, cross correlation and skewness at high redshift 
from a present observational data.

%%%%%%%%%%%%%%%%%%%%%
As mentioned in Sec.\ref{sec: intro}, the recent observation indicates that 
the high redshift galaxies at $z\simeq3$ is largely biased, $b\simeq 6.2$ 
(\cite{P98}, see also \cite{Pea98}). On the other hand, the correlation 
functions of the 
present optical galaxies are one half time larger than that of 
the {\it IRAS} galaxies, which means that the bias parameter for the 
optical galaxies can be evaluated as $b(z=0)=1.5$ if we require no bias 
to the {\it IRAS} galaxies (\cite{H97},\cite{SDYH92}). This suggests that the 
time evolution plays an important role for explaining the difference of 
the bias parameter. The present bias parameter is also evaluated by 
the observation of the higher order statistics for various galaxy survey, 
the Lick catalog, for example. 
Using the prediction from the perturbation theory, Fry examined the 
deterministic non-linear bias and obtained $b=3.5$ and $b_2=14.7$ to 
explain the flat shape of $Q_g$ for the Lick 
catalog (\cite{F94}). This contradicts with the result inferred from 
{\it IRAS} survey. 
%%%%%%%%%%%%%%%%%%%%%%%%

In the stochastic description of the bias effect, taking into account the 
cross correlation between $\dm$ and $\dg$, we can fit the 
bi-spectrum to the Lick catalog so as to be consistent with 
the bias parameter inferred 
from {\it IRAS} galaxy survey. Using the best fit parameters of 
the stochastic bias, we will see 
the evidence of large biasing for high redshift galaxies and evaluate the 
initial skewness at $z=3$. 
%%%%%%%%%%%%%%%%%%%%%%%%
Let us denote $b_0$ and $r_0$ as the initial bias parameter and 
the initial cross correlation set at $z=3$, respectively.  
Providing the present bias parameter as $b=1.5$, the initial 
bias parameter $b_0$ is written in terms of the cross correlation $r_0$: 
\begin{equation}
  b_0=3\left(\sqrt{r_0^2+3}-r_0\right).
  \label{high-redshift-b}
\end{equation}
Substituting the above equation into 
(\ref{linear-t-bias-parameter}), $b(t)$ becomes a 
function of $r_0$. Thus we can know the initial bias 
parameter from the final condition $b(z=0)=1.5$  by varying the $r_0$. 
Fig.3 represents the time reversal evolution of $b$ for the initial 
cross correlation $0\leq r_0 \leq1$. In this inverse problem, we have the 
large bias parameter $b_0=5.20$ when $r_0=0.0$. 

%%%%%%%%%%%%%%%%%%%%%%%%
The unknown parameter $r_0$ can be determined from 
the present bi-spectrum of the Lick catalog. 
The initial amplitude 
and the configuration shape of the bi-spectrum at $z=3$ 
change due to the evolution. Using the fact that 
the spectral index of the Lick catalog becomes 
$n=-1.41$ on large scales (\cite{F94}), 
the present bi-spectrum $Q_g(z=0)$ can be given as a 
function of $r_0$ and $h$. 
In Sec.\ref{subsec: bi-spectrum}, we observed that the bias parameter $b_0$ 
affects the configuration shape of $Q_g$ (for the larger value $b_0$, 
the shape becomes flatter). In the present case, $b_0$ is associated with 
$r_0$ by (\ref{high-redshift-b}). On the other hand, 
the initial skewness parameter 
$h$ can change the amplitude of the bi-spectrum. Therefore, adjusting 
$h$ and $r_0$, 
we can obtain the bi-spectrum $Q_g$ consistent with the Lick catalog. 
%%%%%%%%%%%%%%%%%%%%%%%%

In Fig.4, assuming the Einstein-de Sitter universe,   
the bi-spectrum $Q_g$ at $z=0$ is depicted by varying the initial 
 cross correlation $0\leq r_0 \leq 1$ (solid lines). 
We use the spectral index $n=-1.41$ same as the Lick catalog. 
The dashed line corresponds to the 
bi-spectrum of the Lick catalog with the best fitting parameters 
$b=3.5,~b_2=14.7$ obtained by Fry (\cite{F94}). The figure shows that 
the configuration shape of the Lick catalog coincides with the 
bi-spectrum given by the initial cross correlation $r_0=0.2$. Therefore,  
the bias parameter at high redshift $z=3$ is evaluated $b_0=4.63$ 
from (\ref{high-redshift-b}). In this figure, 
the initial skewness is chosen as $h=6.96$. This result indicates that 
the large-scale galaxy distribution significantly deviates from Gaussian 
distribution and has a large bias parameter. This is consistent 
with the recent observation of Lyman-limit galaxies, which 
has $b\simeq 6.2$ on small scales $\sim7.5$Mpc (\cite{P98}). 
Note that the present skewness evaluated by the best fit parameters is 
$S_{3,g}=4.55$ from (\ref{3-moment}) and (\ref{skewness-g}), 
which also agrees with the observation of the Lick catalog (\cite{G92}).  
%%%%%%%%%%%%%%%%%%%%%%%%

Another explanation of the flat shape of the bi-spectrum in the 
Lick catalog is that the non-linear evolution eliminates the shape 
dependence (\cite{Sc98}). 
If this is true, our tree-level analysis breaks and the 
next-to-leading order perturbations (loop correction) are needed 
(\cite{Sc97}). 
However, the observation on large scales guarantees the quasi non-linear 
evolution. Thus the most appropriate answer would be the bias effect. 
We should remark that the analysis discussed here can also apply to 
an arbitrary high redshift galaxies as well as the galaxies at $z=3$. 
We can predict the bias parameter, cross correlation and skewness at high 
redshift if we obtain the bias parameter, the  
cross correlation and the skewness at present consistent with the 
observation of the bi-spectrum. 
Although this prediction does not take into account 
the formation process of galaxies or morphology of the galaxy type, 
one can say the presence of cross correlation is necessary to lead to 
the consistent 
result with the observational data. Therefore, the stochastic bias is 
essential to describe the galaxy distribution.  
%
%
%
%%%%%%%%%%%%%%%%%%%%%%%%%%%%%%%%%%%%%%%%%%%%%%%%%%%%%%%%%%%%%%%%%%%%%%
%%%%%%%%%%%%%%%%%%%%%%%%%%%%%%%%%%%%%%%%%%%%%%%%%%%%%%%%%%%%%%%%%%%%%%
\section{Conclusion}
\label{sec: summary}
%%%%%%%%%%%%%%%%%%%%%%%%%%%%%%%%%%%%%%%%%%%%%%%%%%%%%%%%%%%%%%%%%%%%%%
%%%%%%%%%%%%%%%%%%%%%%%%%%%%%%%%%%%%%%%%%%%%%%%%%%%%%%%%%%%%%%%%%%%%%%
%
%
In this paper, we have developed the formalism of stochastic bias and 
investigated the quasi non-linear evolution of the galaxy biasing 
under the influence of gravity. After deriving the evolution equations, 
we analyzed the influence of the cross correlation between $\dm$ and $\dg$ 
on the evolution of skewness and bi-spectrum of galaxy distribution. 
We observed that the stochastic description differs from the 
deterministic biasing. The small value of the cross correlation 
gives rise to the rapid growth of skewness and steep shape of the 
bi-spectrum. 
Our main results can be summarized as follows: 
\begin{itemize}
%%%%%%%%%%
\item
 We found the dynamical feature of the stochastic bias 
from the weakly non-linear analysis. 
The gravity enforces the $\dg$-$\dm$ relation to become unity 
and the larger fluctuation accelerates this behavior. 
The small initial cross correlation enhances the 
accerelation and leads to the non-linear $\dg$-$\dm$ relation 
with a negative curvature $b_2<0$.
The result implies that 
the gravity significantly affects the stochastic property 
of the galaxy and the mass. Hence, the dynamical evolution of the 
stochastic bias should be taken into account when 
we interpret the observation of the large scale structure. 
%%%%%%%%%%
\item
 A method for predicting the bias, cross correlation and skewness at 
high redshift was presented. Using the present observation of the 
bi-spectrum and the bias parameter inferred from {\it IRAS} galaxy survey, 
the stochastic property of $\dm$ and $\dg$ at present can be determined. 
The parameters $b$, $r$ and $S_{3,g}$ at high redshift are obtainable 
by the time reversal 
evolution. We solved this inverse problem and predicted the large scale 
bias $b=4.63$ at $z=3$ by fitting the bi-spectrum to the Lick catalog. 
The method is easily applicable to an arbitrary high redshift survey and 
will provide a probe for investigating the clustering property of 
high redshift galaxies.   
\end{itemize}
These results show that the cross 
correlation plays an important role and the consideration of the 
dynamical evolution of the stochastic bias is required to 
explain the observational data. 
Thus, it turns out that the stochastic bias and its dynamics are 
essential to explore the evolution of large scale structure.  

Future sky surveys will provide us a lot of information of 
galaxy distribution. The time evolution of 
the statistical quantities of the large scale distribution will be observed. 
Since the stochasticity of galaxy and mass distribution cannot be removed, 
to use stochastic bias is inevitable for investigating 
the evolution of large scale structure. 
Therefore it is important to 
understand the stochastic property of the galaxy and mass 
distribution by taking into 
account the time evolution. 
The analytic study of the distribution 
function may give us the important information for the measurement of 
cosmological parameter and test of inflation theory. 
As for this direction, we have obtained 
the non-linear $\dm$-$\dg$ relation in $(\dm,~\dg)$-plane 
although the evaluated relation is a kind of expectation value.
The most relevant object for the observation is the peaks of the joint 
probability functional which gives 
the most probable galaxy-mass relation in the measurement of galaxy 
distribution. The primary goal is to clarify the dynamical evolution 
of the joint probability distribution function ${\cal P}[\dm, \dg]$. 
Weakly non-linear analysis using the Edgeworth expansion would be useful 
to study the dynamics of the stochastic bias 
on large scales. Secondary the 
large scale motion of the galaxies should be explored. 
Since the velocity field of the galaxy distribution itself has 
stochasticity, 
it is important to examine the effect of stochastic bias on 
the {\it POTENT} analysis. 
Although we studied the relation between $\dm$ and $\dg$ 
in this paper, our formalism should be extended to include 
the velocity field. 
We will focus on these issues in the future work(\cite{TKS98}). 

%
%
%
%
%
%
%
%%%%%%%%%%%%%%%%%%%%%%%%%%%%%%%%%%%%%%%%%%%%%%%%%%%%%%%%%%%%%%%%%%%%%%
%%%%%%%%%%%%%%%%%%%%%%%%%%%%%%%%%%%%%%%%%%%%%%%%%%%%%%%%%%%%%%%%%%%%%%
%
%
%
\acknowledgments

The authors thank Masaaki Sakagami for helping us to do the numerical 
calculations and useful discussions. They are also grateful to Ofer Lahav 
and Paolo Catelan for useful comments. 
This work is partially supported by Monbusho 
Grant-in-Aid for Scientific Research 10740118.
\clearpage
%%%%%%%%%%%%%%%%%%%%%%%%%%%%%%%%%%%%%%%%%%%%%%%%%%%%%%%%%%%%%%%%%%%%%%
%%%%%%%%%%%%%%%%%%           References         %%%%%%%%%%%%%%%%%%%%%%
%%%%%%%%%%%%%%%%%%%%%%%%%%%%%%%%%%%%%%%%%%%%%%%%%%%%%%%%%%%%%%%%%%%%%%
%
%
%

%
%
%
%
%
\newpage
%
%
%%%%%%%%%%%%%%%%%%%%%%%%%%%%%%%%%%%%%%%%%%%%%%%%%%%%%%%%%%%%%%%%% 
%%%%%%%%%%%%%%%%%%           Table           %%%%%%%%%%%%%%%%%%%%
%%%%%%%%%%%%%%%%%%%%%%%%%%%%%%%%%%%%%%%%%%%%%%%%%%%%%%%%%%%%%%%%% 
%
%
\begin{deluxetable}{llccccccc}
\tablecolumns{9}
\tablecaption{The present parameters $b,~b_2$ and  
$S_{3,g}$ for the various initial conditions given at $z=3$}
\tablewidth{0pc}
\tablehead{
 \colhead{} &  \colhead{} &\multicolumn{3}{c}{Initial parameters} &
 \colhead{} &   \multicolumn{3}{c}{present values} \\
 \colhead{} &  \colhead{} &\multicolumn{3}{c}{$(z=3)$} &
 \colhead{} &   \multicolumn{3}{c}{$(z=0)$} \\
 \cline{3-5}\cline{7-9}\\
 \colhead{$\Omega_0$} & \colhead{$n$} &  
 \colhead{$b_0$} & \colhead{$r_0$} & \colhead{$S_{3,g}$}&\colhead{}& 
 \colhead{$b$} & \colhead{$b_2$} & \colhead{$S_{3,g}$}}
\startdata
 1.0 & -3 &   0.5 & \phs 0.0 & \phs 3.0 && 0.760 & -0.088 & 5.93 \nl
 1.0 & -3 &   1.0 & \phs 0.0 & \phs 3.0 && 0.790 & -0.133 & 5.51 \nl
 1.0 & -3 &   2.0 & \phs 0.0 & \phs 3.0 && 0.901 & -0.245 & 4.48 \nl
 1.0 & -3 &   2.0 & \phs 0.5 & \phs 3.0 && 1.090 & -0.088 & 4.23 \nl
 1.0 & -3 &   2.0 & \phs 1.0 & \phs 3.0 && 1.250 &  0.119 & 4.11 \nl
 1.0 & -1 &   2.0 & \phs 0.5 & \phs 3.0 && 1.090 &  0.033 & 2.71 \nl
 1.0 & -1 &   2.0 & \phs 0.5 & \phs 0.0 && 1.090 & -0.128 & 2.30 \nl
 0.2 & -3 &   2.0 & \phs 0.5 & \phs 3.0 && 1.30 &  -0.132 & 4.49 \nl
\enddata
\end{deluxetable}
\clearpage
%%%%%%%%%%%%%%%%%%%%%%%%%%%%%%%%%%%%%%%%%%%%%%%%%%%%%%%%%%%%%%%%% 
%%%%%%%%%%%%%%%%%%       Figure  Caption     %%%%%%%%%%%%%%%%%%%%
%%%%%%%%%%%%%%%%%%%%%%%%%%%%%%%%%%%%%%%%%%%%%%%%%%%%%%%%%%%%%%%%% 
%
%
%
\section*{Figure Caption}
\begin{description}
%%%%%%%%%%%%%%%%%%%%%%%%%%%%
%
%
%
\item[Fig.1] 
%Evolution of the skewness $S_{3,g}$: 
The time evolutions of the 
skewness $S_{3,g}$ with the specific parameters $\Omega_0=1, n=-3, h=3.0$ 
are plotted as a function of expansion factor $a$. 
The initial conditions for each lines are given at $a=1$ as follows:
$b_0=2.0, r_0=1.0$({\it solid line}),  
$b_0=2.0, r_0=0.5$({\it long-dashed line}),  
$b_0=1.0, r_0=0.0$({\it short-dashed line}),  
$b_0=0.5, r_0=0.0$({\it dotted line}). 
 These lines asymptotically approach $34/7$, which corresponds to the 
skewness of the total mass distribution. 
%
%
%
%%%%%%%%%%%%%%%%%%%%%%%%%
\item[Fig.2] 
The evolved results of the bi-spectrum amplitude at $z=0$ 
with the spectral index $n=-3$. We plotted $Q_g$ as a function of 
$\theta=\cos^{-1}(\hat{\k}_1\cdot\hat{\k}_2)$ setting $k_1/k_2=2$. 
Each line has the same initial skewness $h=3.0$ which are given at $z=3$ 
and the universe is assumed to be a Einstein-de Sitter universe($\Omega_0=1$): 
(a). Variation of $Q_g(\theta)$ when the initial bias parameter is fixed 
($b_0=2.0$). The initial values of the cross correlation 
for each solid lines are 
$r_0=0.0,~0.25, 0.5,~0.75$ and $1.0$({\it from top to bottom}). 
We see that the smaller cross correlation makes the configuration 
shape of $Q_g$ steeper. 
(b). Shape and amplitude dependences of $Q_g(\theta)$ when the initial cross 
correlation is fixed ($r_0=0.5$). 
The results of the bi-spectrum have the initial bias parameters given by 
$b_0=0.0,~0.5, 1.0,~2.0$ and $3.0$ ({\it from top to bottom}). 
Compared with both figures, influence of the cross correlation is weaker than 
that of the bias parameter.
%
%
%
%%%%%%%%%%%%%%%%%%%%%%%%%
\item[Fig.3] 
%Evolution of the bias parameter $b$: 
The time reversal evolution of the bias parameters 
(\ref{linear-t-bias-parameter}) when the 
present ($z=0$) bias parameter is given by $b=1.5$. 
The horizontal axis denotes the redshift parameter $z$. 
The results are shown for the cross correlation 
$r_0=0.0, 
0.2, 0.4, 0.6, 0.8$ and $1.0$ ({\it from top to bottom}), which are 
the values set at $z=3$. The figure shows that 
the bias parameter takes the values from $3$ to $5.2$ at $z=3$.
%%%%%%%%%%%%%%%%%%%%%%%%%%%%
\item[Fig.4] 
The bi-spectrum $Q_g$ by setting the initial 
conditions at $z=3$, all of which have the same bias parameter $b=1.5$ 
at present (solid lines). We use the spectral index $n=-1.41$ 
same as the Lick catalog. 
Each lines denote the present value of bi-spectrum 
with the initial cross correlation 
$r_0=0.0$, $0.2, 0.4, 0.6, 0.8$ and $1.0$({\it from top to bottom}), 
respectively. These lines have the same initial skewness $h=6.96$. 
The dashed line corresponds to the bi-spectrum of the 
Lick catalog constructed by the time independent non-linear bias 
(\cite{F94}). We can fit the bias parameter to the Lick catalog if the 
initial cross correlation is chosen as $r_0=0.2$, which leads to 
the initial bias parameter $b_0=4.63$. This indicates that the 
large-scale distribution of high redshift galaxies is largely biased 
and has large skewness.
 \end{description}
%
%
%
%%%%%%%%%%%%%%%%%%%%%%%%%%%%%%%%%%%%%%%%%%%%%%%%%%%%%%%%%%%%%%%%% 
%%%%%%%%%%%%%%%%%%            Figures        %%%%%%%%%%%%%%%%%%%%
%%%%%%%%%%%%%%%%%%%%%%%%%%%%%%%%%%%%%%%%%%%%%%%%%%%%%%%%%%%%%%%%% 
%
%
%
%
%
%
\newpage
\epsfxsize=\hsize
\epsfbox{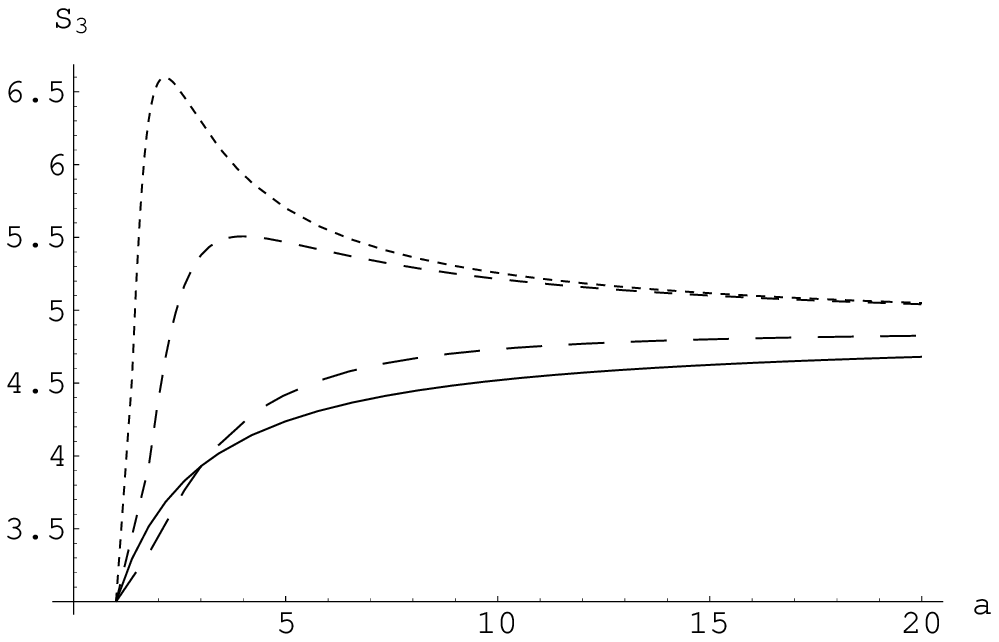}     
\vspace{2.5cm}
\begin{center}
{\large Fig.1}
\end{center}

\epsfxsize=\hsize
\epsfbox{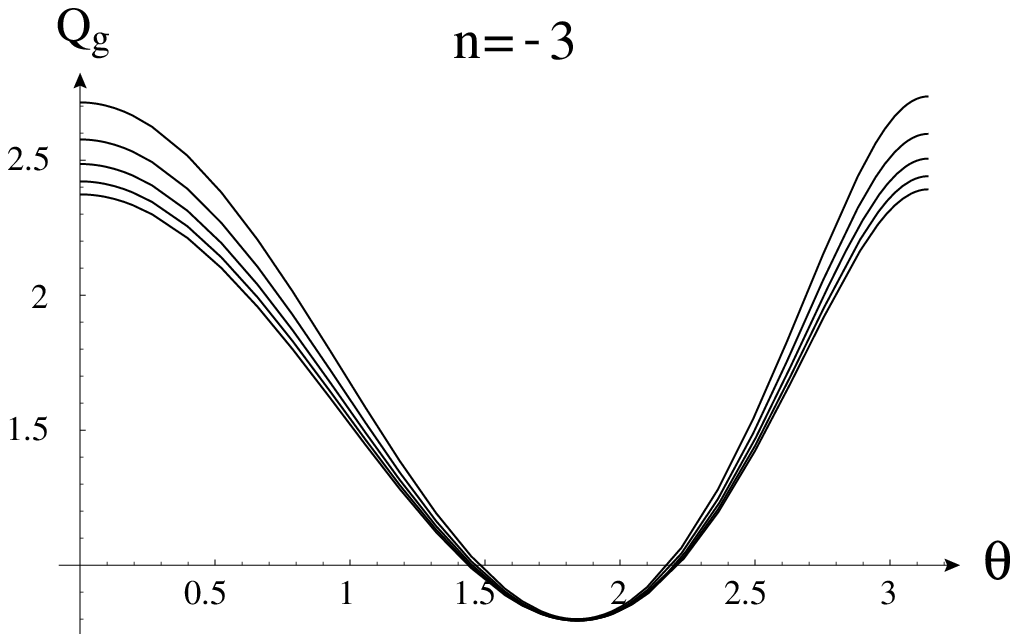}     
\vspace{2.5cm}
\begin{center}
{\large Fig.2a}
\end{center}

\epsfxsize=\hsize
\epsfbox{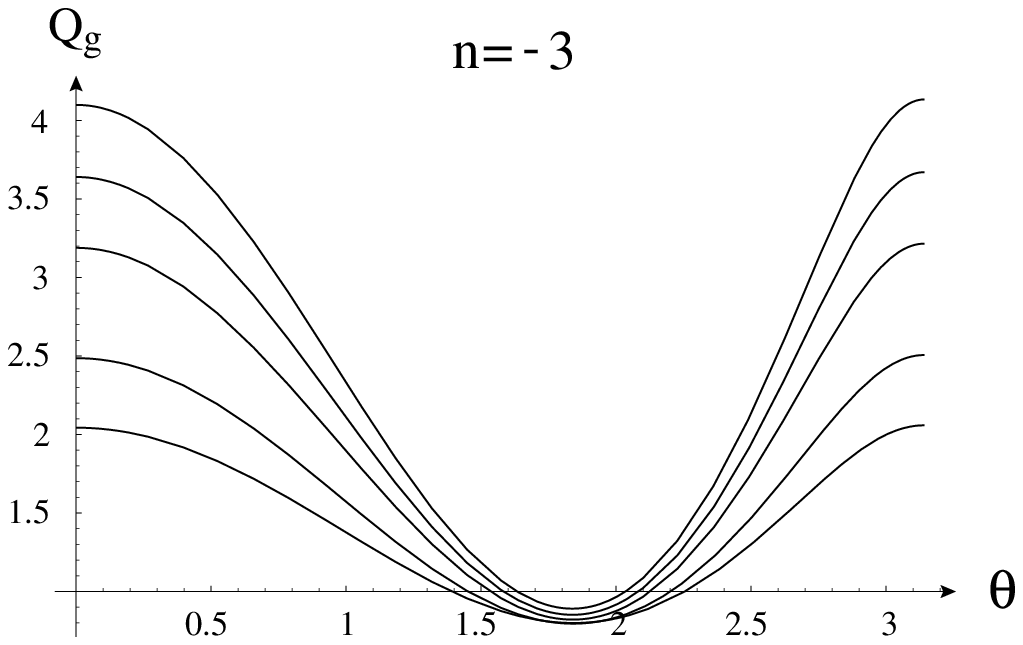}     
\vspace{2.5cm}
\begin{center}
{\large Fig.2b}
\end{center}

\epsfxsize=\hsize
\epsfbox{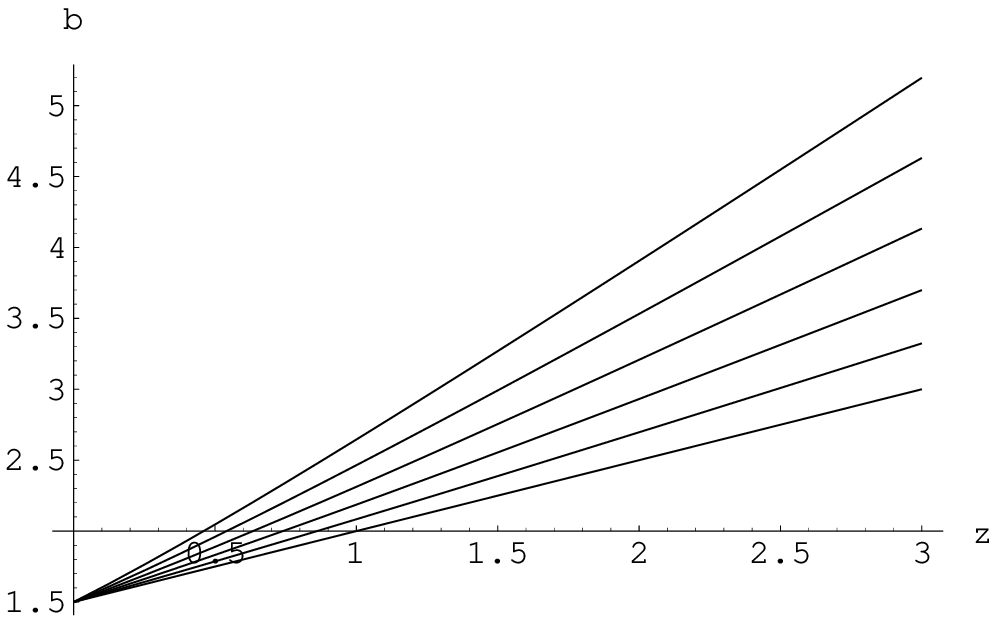}     
\vspace{2.5cm}
\begin{center}
{\large Fig.3}
\end{center}

\epsfxsize=\hsize
\epsfbox{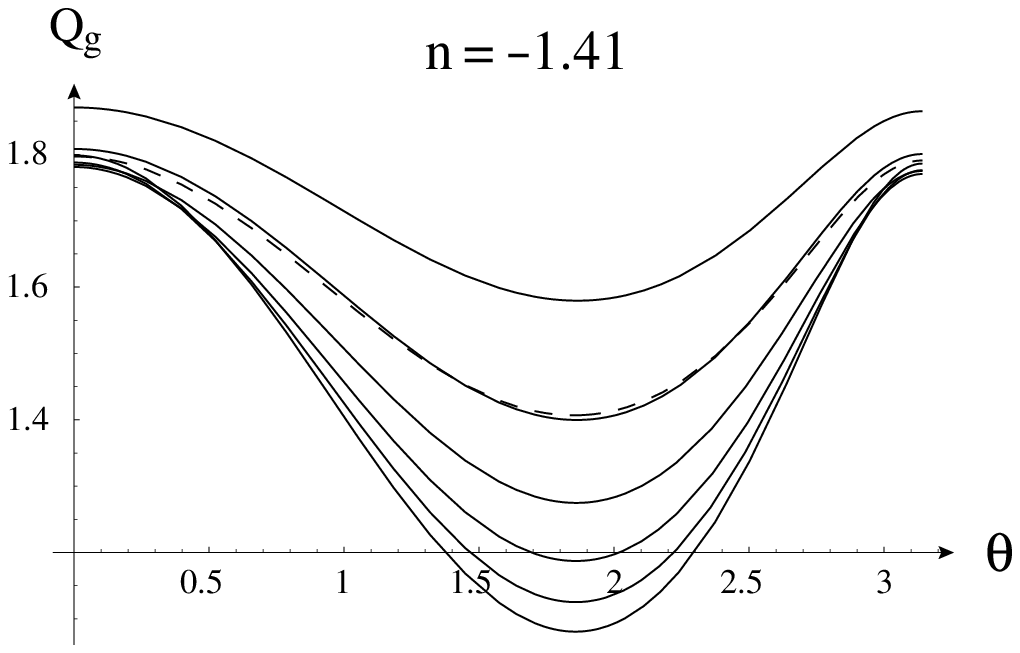}     
\vspace{2.5cm}
\begin{center}
{\large Fig.4}
\end{center}

\begin{thebibliography}{99}
%%%%%%%
\bibitem[Bernardeau 1994a]{B94a} Bernardeau, F., 1994, ApJ, 427, 51.
%%%%%%%
\bibitem[Bernardeau 1994b]{B94b} Bernardeau, F., 1994, ApJ, 433, 1.
%%%%%%%
\bibitem[Cen \& Ostriker 1992]{CO97} Cen, R., \& Ostriker, J.P., 
1992, ApJ, 399, L113.
%%%%%%%
\bibitem[Catelan {\it et al}.1998a]{catelan-a} Catelan, P., Lucchin, F., Matarrese, S., 
\& Porciani, C., 1998, {\it MNRAS},  297, 692.
%%%%%%%
\bibitem[Catelan {\it et al}.1998b]{catelan-b} Catelan, P., Matarrese, S., 
\& Porciani, C., 1998, ApJL, in press.
%%%%%%%
\bibitem[Dekel 1997]{Dekel} Dekel, A., 1997, astro-ph/ 9705033.
%%%%%%
\bibitem[Dekel and Lahav 1998]{DL98} 
  Dekel, A., \& Lahav, O., 1998, astro-ph/ 9806193.
%%%%%%%
\bibitem[Frieman \& Gazta\~naga 1994]{FG94} 
Frieman, J.A., \& Gazta\~naga, E., 1994,   ApJ, 425, 392.
%%%%%%%
\bibitem[Fry 1984]{F84} Fry, J.N., 1984, ApJ, 279, 499.
%%%%%%%
\bibitem[Fry \& Gazta\~naga 1993]{FG93} Fry, J.N., \& Gazta\~naga, E., 
  1993, ApJ, 413, 447.
%%%%%%%
\bibitem[Fry 1994]{F94} Fry, J.N., 1994, Phys.Rev.Lett, 73, 215.
%%%%%%%
\bibitem[Fry 1996]{F96} Fry, J.N., 1996, ApJ, 461, L65.
%%%%%%%
\bibitem[Gazta\~naga 1992]{G92} Gazta\~naga, E., 1992, ApJ, 398, L17.
%%%%%%%
\bibitem[Gazta\~naga \& Frieman 1994]{GF94} 
Gazta\~naga, E.,\& Frieman, J.A., 1994,   ApJ, 437, L13.
%%%%%%%
\bibitem[Hamilton 1997]{H97} Hamilton, A.J.S., 1997, astro-ph/9708102.
%%%%%%%
\bibitem[Juszkiewicz {\it et al}.1995]{JWACB95} Juszkiewicz, R., Weinberg, D.H., 
Amsterdamski, P., Chodorowski, M., \& Bouchet, F.R., 1995, ApJ, 442, 39.
%%%%%%%
\bibitem[Kaiser 1984]{K84} Kaiser, N., 1984, ApJ, 284, L9.
%%%%%%%
\bibitem[Lahav 1996]{Lahav} Lahav, O., 1996, astro-ph/ 9611093.
%%%%%%%
\bibitem[Linde 1990]{Linde} Linde, A.D., 1990, 
{\it Inflation and Quantum Cosmology} (Academic Press, Inc.)
%%%%%%%
\bibitem[Mann {\it et al}. 1997]{Mann97} Mann, R.G., Peacock, J.A., 
\& Heavens, A.F.,
 1997, astro-ph/ 9708031.
%%%%%%%
\bibitem[Matarrese {\it et al}.1986]{MLB86} 
  Matarrese, S., Lucchin, F., \& Bonometto, S.A., 1986, ApJ, 310, L21. 
%%%%%%%
\bibitem[Mo \& White 1996]{MW96} Mo, H.J., \& White, S.D.M., 
1996, {\it MNRAS}, 282, 347.
%%%%%%%
\bibitem[Porciani {\it et al}.1998]{Porciani} Porciani, C., Matarrese, S., 
Lucchin, F., \& Catalen, P., 1998, {\it MNRAS}, in press.
%%%%%%%
\bibitem[Peacock 1998]{P98} Peacock, J.A., 1998, astro-ph/ 9805208. 
%%%%%%%
\bibitem[Peacock {\it et al}. 1998]{Pea98} 
  Peacock, J.A., Jimenez, R., Dunlop, J.S., Waddington, I., Spinrad, H., 
  Stern, D., Dey, A., \& Windhorst, R.A., 1998, MNRAS, 296, 1089. 
%%%%%%%
\bibitem[Peebles 1980]{Peebles} Peebles, P.J.E., 1980, 
{\it The Large-Scale Structure of the Universe} (Princeton U.P., Princeton)
%%%%%%%
\bibitem[Pen 1997]{Pen97} Pen, U., 1997, astro-ph/ 9711180.
%%%%%%
\bibitem[Scoccimarro 1997]{Sc97} Scoccimarro, R., 1997, ApJ, 487, 1.
%%%%%%
\bibitem[Scoccimarro {\it et al}. 1998]{Sc98} Scoccimarro, R., 
Colombi, S, Fry, J.N., Frieman, J.A., Hivon, E., \& Melott, A., 
1998, ApJ, 496, 586.
%%%%%%
\bibitem[Strauss {\it et al.} 1992]{SDYH92} Strauss, M.A., 
Davis, M., Yahil, A., \& Huchra, J.P. 1992, ApJ, 385, 421.
%%%%%%
\bibitem[Taruya, Koyama \& Soda 1998]{TKS98} 
  Taruya, A., Koyama, K., \& Soda, J., in preparation.
%%%%%%%
\bibitem[Tegmark \& Peebles 1998]{TP98} Tegmark, M., \& Peebles, P.J.E., 
1998, ApJ, 500, L79; astro-ph/ 9804067. 
\end{thebibliography}
\end{document}